\documentclass[a4paper,11pt]{article}
\usepackage{pos}
\usepackage{hyperref}

\title{Baryon-Electric Charge Correlations and Chemical Potentials as Probes of Magnetized QCD}
\ShortTitle{\rm{BQ} Correlations and Chemical Potentials as Probes of Magnetized QCD}

\author{Heng-Tong Ding}
\author*{Jin-Biao Gu}
\author{Arpith Kumar}
\author{Sheng-Tai Li}
\author{Jun-Hong Liu}

\affiliation{Key Laboratory of Quark and Lepton Physics (MOE) and Institute of Particle Physics,\\
Central China Normal University, Wuhan 430079, China}
\emailAdd{jinbiaogu@mails.ccnu.edu.cn}

\abstract{We present the first lattice QCD results of quadratic fluctuations and correlations of conserved charges in (2+1)-flavor lattice QCD in the presence of a background magnetic field. The simulations were performed using the Highly Improved Staggered Quarks with physical pion mass $m_\pi$ = 135 MeV on $N_\tau=8$ and 12 lattices. We find that the correlation between net baryon number and electric charge, denoted as $\chi^{\rm BQ}_{11} $, can serve as a magnetometer of QCD. At pseudocritical temperatures ($T_{pc}$) the $\chi^{\rm BQ}_{11}$ starts to increase rapidly with magnetic field strength $eB \gtrsim 2M^2_{\pi}$ and by a factor 2 at $eB\simeq 8 M^2_{\pi}$. By comparing with the hadron resonance gas model, we find that the $eB$ dependence of $\chi^{\rm BQ}_{11}$ is mainly due to the doubly charged $\Delta$(1232) baryon. Although the doubly charged $\Delta$(1232) could not be detected experimentally, its decay products, protons and pions, retain the $eB$ dependence of $\Delta$(1232)'s contribution to $\chi^{\rm BQ}_{11}$. Furthermore, the ratio of electric charge chemical potential to baryon chemical potential, $\mu_{\rm Q}/\mu_{\rm B}$, shows significant dependence on the magnetic field strength and varies with the ratio of electric charge to baryon number in the colliding nuclei in heavy ion collisions. These results provide baselines for effective theory and model studies, and both $\chi^{\rm BQ}_{11}$ and $\mu_{\rm Q}/\mu_{\rm B}$ could be useful probes for the detection of magnetic fields in relativistic heavy ion collision experiments as compared with corresponding results from the hadron resonance gas model.
}

\FullConference{The 41st International Symposium on Lattice Field Theory (LATTICE2024)\\
 28 July - 3 August 2024\\
Liverpool, UK\\
}

\begin{document}

\maketitle

\section{Introduction}

Strong magnetic fields, conjectured to exist in relativistic heavy-ion collisions  \cite{Kharzeev:2007jp,Deng:2012pc,Skokov:2009qp}, magnetars \cite{Enqvist:1993np}, and the early universe \cite{Vachaspati:1991nm}, play a pivotal role in governing the behavior of quantum chromodynamics (QCD) matter. Model estimates suggest that the initial magnetic fields produced in non-central heavy-ion collisions at RHIC and the LHC may reach magnitudes of $eB\sim5~M_\pi^2$ and $eB\sim70~M_\pi^2$, respectively \cite{Deng:2012pc,Skokov:2009qp}. Such intense fields are expected to significantly modify QCD interactions and leave detectable signatures on final-state observables \cite{Kharzeev:2012ph}.

The most prominent effect introduced by strong magnetic fields is the chiral magnetic effect (CME)\cite{Fukushima:2008xe}, which has attracted intense theoretical and experimental interest \cite{Kharzeev:2020jxw}. The key to the CME lies in the strength and longevity of the magnetic field. However, a critical unresolved question is whether the transient field produced in heavy-ion collisions persists long enough to imprint measurable effects on the final-state particles. While first-principles lattice QCD studies have identified observables exhibiting strong dependence on magnetic fields, such as the splitting between $u$ and $d$ quark chiral condensates under external magnetic fields \cite{Ding:2020hxw}, these quantities remain experimentally inaccessible. 

Fluctuations of and correlations among net baryon number (B), electric charge (Q), and strangeness (S) have long served as powerful tools to probe the QCD phase structure\cite{Ding:2015ona,Fu:2022gou,Luo:2017faz,Pandav:2022xxx}. However, studies of these fluctuations under the presence of magnetic fields remain sparse and largely confined to effective models, like hadron resonance gas (HRG) \cite{Fukushima:2016vix,Ferreira:2018pux,Bhattacharyya:2015pra,Kadam:2019rzo}, Polyakov-Nambu-Jona-Lasinio model (PNJL) \cite{Fu:2013ica}. First-principles lattice QCD calculations are essential to establish model-independent benchmarks. The lattice QCD investigation of fluctuations of conserved charges in external magnetic fields has been conducted with an unphysical pion mass ($M_\pi\simeq 220~\rm MeV$) and a single lattice spacing in $N_f=2+1$ QCD \cite{Ding:2021cwv} , and more recently at the physical point in $N_f=2+1+1$ QCD on $N_\tau=8$ lattices ~\cite{Borsanyi:2023buy}.

In this work, we present the first (2+1)-flavor lattice QCD investigation of quadratic fluctuations and charge correlations in a background magnetic field with the physical pion mass. Our results establish model-independent benchmarks and reveal a significant enhancement of both the baryon electric charge correlations, $\chi^{\rm BQ}_{11}$ and the ratio of electric charge chemical potential over baryon chemical potential, ${{\mu}_{\rm Q}}/{\mu}_{\rm B}$ in the presence of a magnetic field. These quantities are promising candidates for probing magnetic fields in relativistic heavy-ion collisions. This proceeding is based on the work presented in Ref. \cite{Ding:2022uwj,Ding:2023bft}.

\section{Fluctuations of conserved charges and hadron resonance gas model}

The fluctuations and correlations of conserved charge can be obtained by taking the derivatives of pressure with respect to the chemical potentials from lattice calculation evaluated at vanishing chemical potentials,
\begin{equation}
    \chi_{i j k}^{\mathrm{BQS}}=\left.\frac{\partial^{i+j+k} p / T^4}{\partial\hat{\mu}_{\mathrm{B}} ^i \partial\hat{\mu}_{\mathrm{Q}} ^j \partial\hat{\mu}_{\mathrm{S}} ^k}\right|_{\hat{\mu}_{\mathrm{B}, \mathrm{Q}, \mathrm{S}}=0},
    \label{eq:def_chi}
\end{equation}
where $\hat{\mu}_X\equiv \mu_X/T$ with $X=$ \{B, Q, S\} and $p=\frac{T}{V} \ln Z\left(V, T, \mu_{\rm B}, \mu_{\rm Q}, \mu_{\rm S}, e B\right)$ denotes the pressure of the magnetized thermal medium. 

In the hadron resonance gas model, the pressure arising from charged and neutral particles in the presence of a magnetic field can be expressed as follows\cite{HotQCD:2012fhj,Ding:2021cwv}
\begin{equation}
    \frac{p_n}{T^4}=\frac{g_i m_i^2}{2(\pi T)^2} \sum_{n=1}^{\infty}( \pm 1)^{n+1} \frac{e^{n \mu_i / T}}{n^2} \mathrm{~K}_2\left(\frac{n m_i}{T}\right),
    \label{eq:HRG_neutral_p}
\end{equation}
\begin{equation}
    \frac{p_c}{T^4}=\frac{\left|q_i\right| B}{2 \pi^2 T^3} \sum_{s_z=-s_i}^{s_i} \sum_{l=0}^{\infty} \varepsilon_l \sum_{n=1}^{\infty}( \pm 1)^{n+1} \frac{e^{n \mu_i / T}}{n} \mathrm{~K}_1\left(\frac{n \varepsilon_l}{T}\right),
    \label{eq:HRG_charged_p}
\end{equation}
where $q_i$, $m_i$, $s_i$ and $g_i$ are the charge, mass, spin and degeneracy factor of the particle $i$. $B$ is the magnitude of the magnetic field pointing along the z direction. $\varepsilon_l=\sqrt{m_i^{2}+2\left|q_i\right| B\left(l+1 / 2-s_{z}\right)}$ are the energy levels of charged particles with $l$ denotes the Landau level. $n$ is the sum index in the Taylor expansion series. $\rm K_2$ and $\rm K_1$ are the second-order and first-order modified Bessel functions of the second kind, respectively. The “$+$” in the “$\pm$” corresponds to mesons (where $s_{i}$ is an integer) while the “$-$” corresponds to baryons (where $s_{i}$ is a half-integer).

The quadratic fluctuations of and correlations among conserved charge arising from can be expressed by,
\begin{equation}
    \begin{aligned}
    &{\chi}_{2}^{X}=\frac{B}{2 \pi^{2} T^3} \sum_{i}\left|q_{i}\right| X_{i}^{2} \sum_{s_{z}=-s_{i}}^{s_{i}} \sum_{l=0}^{\infty} f\left(\varepsilon_{0}\right) \,,\\
    &{\chi}_{11}^{X Y}=\frac{B}{2 \pi^{2} T^3} \sum_{i}\left|q_{i}\right| X_{i} Y_{i} \sum_{s_{z}=-s_{i}}^{s_{i}} \sum_{l=0}^{\infty} f\left(\varepsilon_{0}\right)\,,
    \end{aligned}
    \label{eq:HRG_sus}
\end{equation}
where $f\left(\varepsilon_{0}\right)=\varepsilon_{0} \sum_{n=1}^{\infty}(\pm 1)^{n+1} n \mathrm{~K}_{1}\left(\frac{n \varepsilon_{0}}{T}\right)$ and $X_i,Y_i=$ $\mathrm{B}_i, \mathrm{Q}_i, \mathrm{S}_i$ carried by hadron $i$. In this work, we focus on the computation of quadratic fluctuations and correlations, i.e. $i+j+k=2$.

\section{Lattice setup}

In this work, we adopt the highly improved staggered quarks (HISQ) \cite{Follana:2006rc} and a tree-level improved Symanzik gauge action to perform lattice simulations of $N_f = 2+1$ QCD in external magnetic fields \cite{Ding:2021cwv}. We primarily use the results from $32^3\times8$ and $48^3 \times 12$ lattices to perform continuum estimate. To validate the reliability of our continuum estimate results, we have conducted additional measurements at a specific temperature and magnetic field value on $64^3 \times 16$ lattice. The constant magnetic field is introduced along the $z$-direction and is described by a fixed factor $u_\mu(n)$ of the $U(1)$ field, which defined within the Landau gauge framework \cite{Bali:2011qj,AlHashimi:2008hr},
\begin{equation}
    \begin{aligned}
    & u_x\left(n_x, n_y, n_z, n_\tau\right)= \begin{cases}\exp \left[-i a^2 q B N_x n_y\right] & \left(n_x=N_x-1\right) \\
    1 & \text { (otherwise) }\end{cases} \\
    & u_y\left(n_x, n_y, n_z, n_\tau\right)=\exp \left[i a^2 q B n_x\right], \\
    & u_z\left(n_x, n_y, n_z, n_\tau\right)=u_t\left(n_x, n_y, n_z, n_\tau\right)=1 ,
    \end{aligned}
\end{equation}

Unlike our previous studies \cite{Ding:2021cwv}, the quark masses are now fixed to their physical values, corresponding to a pion mass of $M_\pi=135$ MeV in the absence of a magnetic field. Our temperature range is centered around $T_{pc}$, spanning from $0.9T_{pc}$ to $1.1T_{pc}$. To mimic the intense magnetic fields generated during the initial stages of relativistic heavy-ion collisions, we implemented six distinct values of magnetic field strengths $eB$ up to $8 M^2_\pi$.

\section{Results and discussions}

For obtaining the continuum estimate, we perform a joint fit using a linear extrapolation in terms of \(1/N_\tau^2\),
\begin{equation}
    \mathcal{O}\left(T,eB, N_{\tau}\right)=\mathcal{O}(T,eB)+\frac{c}{N_{\tau}^{2}},
    \label{eqn:cont_est_linear}
\end{equation}
where $\mathcal{O}(T,eB)$ is the final continuum estimate. 

\begin{figure}[!htp]
    \centering
    \includegraphics[width=0.4\textwidth]{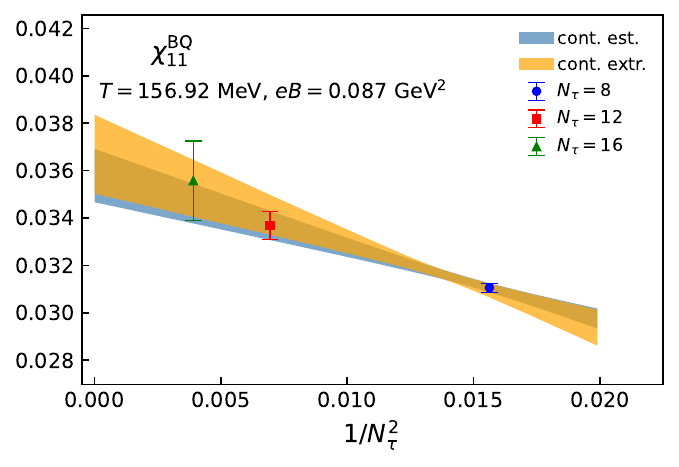}
    \includegraphics[width=0.4\textwidth]{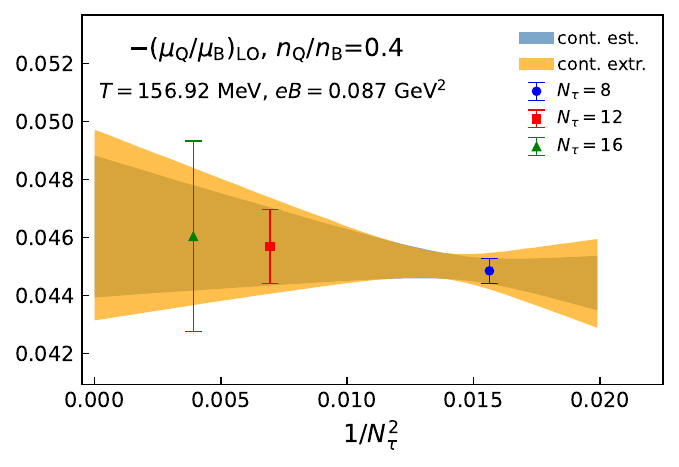}
    \caption{Continuum estimate (blue band) and extrapolation (yellow band) result of $\chi^{\rm{B} \rm{Q}}_{11}$ (left) and $-\left(\mu_{\rm{Q}}/\mu_{\rm{B}}\right)_{\rm LO}$ with $n_{\rm{Q}}/n_{\rm{B}}=0.4$  (right) at $T=156.92$ MeV at $eB=0.087~\text{GeV}^2$. The points for $N_\tau=8$ and 12 are interpolated results while the point for $N_\tau=16$ denotes the lattice data.}
    \label{fig:cont_BQ_q1_eB}
\end{figure}
The left panel of \autoref{fig:cont_BQ_q1_eB} shows the baryon electric charge correlation as a function of $1/N_\tau^2$ at specific temperature and magnetic field. We compare the continuum estimates obtained using $N_\tau = \{8, 12\}$ with the continuum extrapolation derived from $N_\tau = \{8, 12, 16\}$, and the systematic uncertainties between two methods are within the statistical uncertainties. The right panel presents analogous results for $\left(\mu_{\rm{Q}}/\mu_{\rm{B}}\right)_{\rm LO}$, a composite observable derived from six quadratic conserved charge fluctuations and correlations. These consistencies demonstrate the reliability of our continuum estimate method, even when applied to complex observables.

\begin{figure}[!htp]
    \centering
    \includegraphics[width=0.32\textwidth]{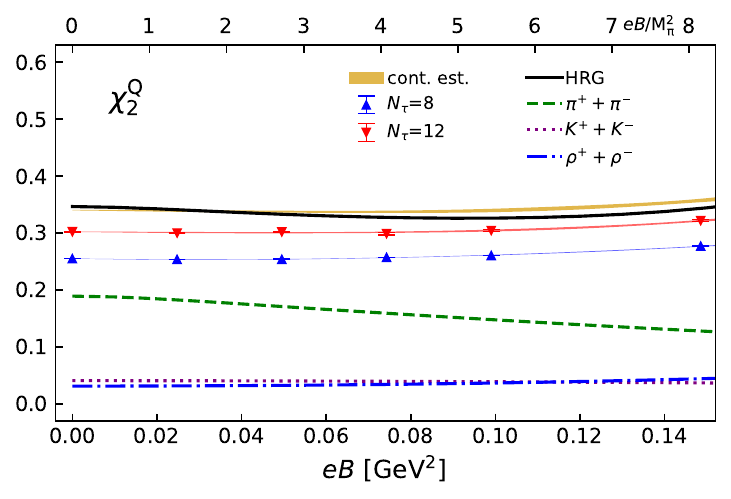}
    \includegraphics[width=0.32\textwidth]{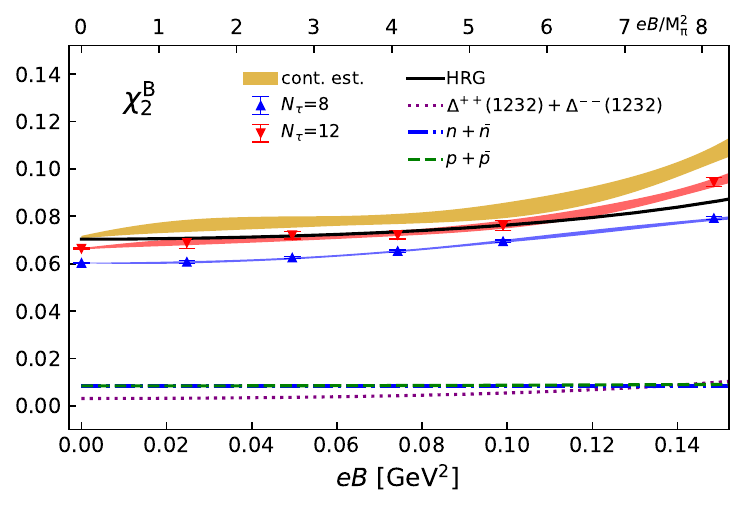}
    \includegraphics[width=0.32\textwidth]{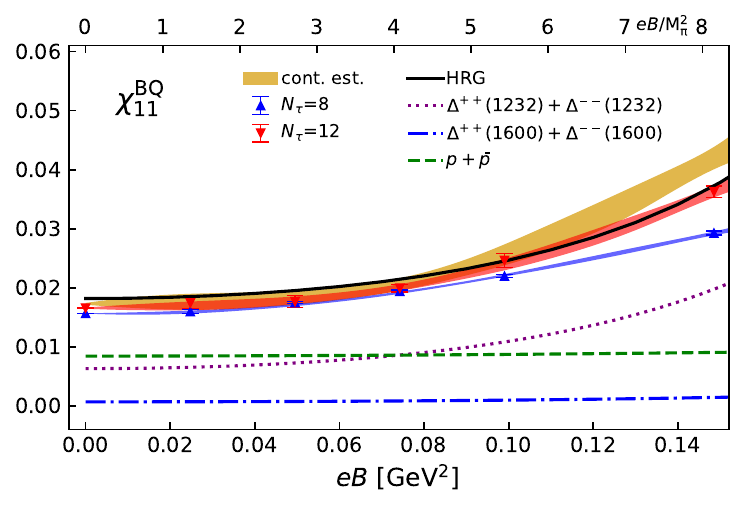}
    \caption{The yellow band represents the continuum estimates  of $\chi^{\rm Q}_{2}$, $\chi^{\rm B}_{2}$ and $\chi^{\rm BQ}_{11}$ at non vanishing magnetic field with $T =145$ MeV. The black solid line and broken line shows the total contribution and contributions from individual hadrons to $\chi^{\rm Q}_{2}$, $\chi^{\rm B}_{2}$ and $\chi^{\rm BQ}_{11}$ obtained from the HRG model.}
    \label{fig:T145_lqcd_hrg}
\end{figure}
\autoref{fig:T145_lqcd_hrg} shows the results of $\chi^{\rm Q}_{2}$, $\chi^{\rm B}_{2}$ and $\chi^{\rm BQ}_{11}$ as a function of magnetic field strength obtained from lattice QCD computation and HRG model at $T =145$ MeV. The color bands in figure are obtained from B-spline interpolation on $T$-$eB$ plane using data from various $N_b$ and $T$ values. Further interpolation details are provided in Ref. \cite{Ding:2023bft}. The lattice continuum estimates are obtained from the $N_\tau=8$ and 12. We find $\chi^{\rm Q}_{2}$ almost independent of $eB$. From the perspective of the HRG model, this can be attributed to the counterbalancing of different hadronic contributions. With increasing of the magnetic field the contribution from half-spin meson, such as pions, decreases, while that from higher-spin hadrons increases.  $\chi^{\rm B}_{2}$ increase with $eB$, but the HRG model fails to capture this trend due to its assumption that the energies of neutral particles are not affected by magnetic fields. $\chi^{\rm BQ}_{11}$ exhibits strikingly sensitivity to the magnetic field, reaching 2.4 times its value at vanishing magnetic field at $eB \simeq 8 M_\pi^2$. Such a remarkable enhancement suggests its potential utility as a magnetometer in QCD. With the aid of the HRG model, we can discern that this significant increase is predominantly driven by the contributions from $\Delta^{++}(1232)$ ($\Delta^{--}(1232)$). However, doubly charged $\Delta$ baryons undergo strong decay and rapidly decay to stable particles such as protons and pions, which implies that they cannot be measured directly in heavy-ion collision (HIC) experiments.

Advantageously, Ref. \cite{Bellwied:2019pxh} provides a method for reconstructing the contributions of resonances using stable particles within the HRG framework. 
Here we employ $\pi$, $K$ and $p$ to construct the proxies for the resonance contributions \cite{STAR:2019ans}. The proxy for $\chi^{\rm Q}_{2}$ and $\chi^{\rm BQ}_{11}$ are defined follows,
\begin{equation}
    \begin{aligned}
       &\sigma_{Q^{\rm PID},p}^{1,1}= \sum_R\left(P_{R \rightarrow \tilde{p}}\right)\left(P_{R \rightarrow Q^{\rm PID}}\right) I_R^{\rm BQ} + I_{\tilde{p}}^{\rm BQ} ,\\
        &\sigma_{Q^{\rm PID}}^{2}= \sum_R\left(P_{R \rightarrow Q^{\rm PID}}\right)^2 I_R^{\rm Q} + I_{Q^{\rm PID}}^{\rm Q} ,
    \end{aligned}
    \label{eq:proxy}
\end{equation}
where $\tilde{j} = j-\bar{j}$ represents the net quantum number of particle species $j$,with $P_{R \rightarrow \tilde{j}} = P_{R \rightarrow j} - P_{R \rightarrow \bar{j}}$. $P_{R \rightarrow j}=\sum_\alpha {N}_{R \rightarrow j}^\alpha n_{j, \alpha}^R$ giving the average number of particle $j$ produced by each particle $R$ after the entire decay chain.  $\alpha$ is a decay channel from $R$ to $j$, ${N}_{R \rightarrow j}^\alpha$ is the branching ratio, and $n_{j, \alpha}^R$ is the number of stable particle $j$ produced from this decay channel.  And $Q^{\rm PID}=\tilde{p}+\tilde{K}+\tilde{\pi}$ represents the summation of contributions from net proton, pion and kaon. We assumed that the branching ratios are independent of the magnetic field and we used all decay modes listed in the Particle Data Group \cite{ParticleDataGroup:2020ssz}. The $I_R^{\rm BQ}$ and $I_R^{\rm Q}$ stands for the contribution from the particle $R$ to $\chi_{11}^{\rm BQ}$ and $\chi_{2}^{\rm Q}$, respectively, where $\chi_{11}^{\rm BQ}\equiv\sum_i I_{i}^{\rm BQ}$ and $\chi_{2}^{\rm Q}\equiv\sum_i I_{i}^{\rm Q}$.

\begin{figure}[!htp]
    \centering
    \includegraphics[width=0.4\textwidth]{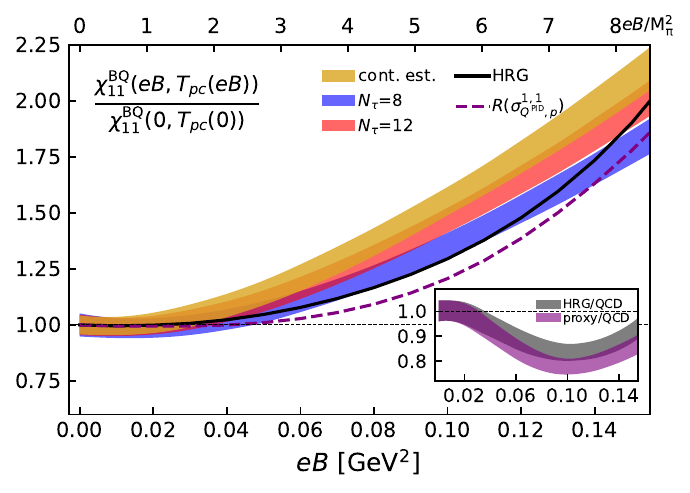}
    \includegraphics[width=0.4\textwidth]{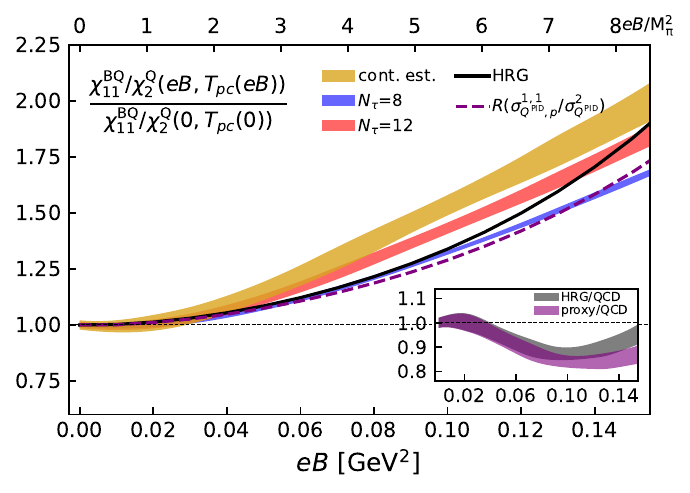}
    \caption{The $R_{cp}$-like ratios of $\chi^{\rm BQ}_{11}$ and $\chi^{\rm BQ}_{11}/\chi^{\rm Q}_{2}$ from lattice QCD to their corresponding values at vanishing magnetic fields (yellow band) along the transition line. The results from HRG (black solid line) and proxy (purple dashed line) are also shown. The insets show the ratios of results calculated from HRG and proxy to lattice computation.}
    \label{fig:Rcp}
\end{figure}

The pseudocritical temperature $T_{pc}$($\sim 156~\rm MeV$) is close to the chemical freeze out temperature in HIC experiments \cite{HotQCD:2018pds,ALICE:2017jmf}. In this context, weak magnetic fields correspond to the central collisions, while strong magnetic fields correspond to the peripheral collisions. Therefore, the ratio of an observable at non-zero magnetic field to its value at vanishing magnetic field defines a $R_{cp}$-like observable. In \autoref{fig:Rcp}, we present $R_{cp}$-like ratios for $\chi^{\rm BQ}_{11}$ and $\chi^{\rm BQ}_{11}/\chi^{\rm Q}_{2}$ along the transition line. It can be observed that the ratio of $\chi^{\rm BQ}_{11}$ exhibits significant dependence on the magnetic field even at higher temperatures. While both the HRG model and the proxy can reproduce the lattice QCD results only within a limited range of magnetic field strengths,$\sim M_\pi^2$, the proxy retains the significant magnetic field dependence feature of $\chi^{\rm BQ}_{11}$.  In the context of HIC experiments, double-ratio observables are commonly utilized to mitigate volume-dependent effects~\cite{STAR:2019ans}. Notably, the proxy for $\chi^{\rm BQ}_{11}/\chi^{\rm Q}_{2}$ mirrors the behavior of $\chi^{\rm BQ}_{11}$, suggesting its viability as a probe for magnetic field effects~\footnote{In the current work, the kinematic cuts in the proxy have not yet been adopted, and their impact will be investigated in future studies~\cite{2ndOrderFluct}.}.  

The baryon chemical potential ($\mu_{\rm B}$) and the electric charge chemical potential ($\mu_{\rm Q}$) can be extracted via thermal statistical fits to the particle yields in experiments\cite{ALICE:2017jmf,Braun-Munzinger:2003pwq,STAR:2017sal}, while their ratio ${{\mu}_{\rm Q}}/{\mu}_{\rm B}$ is accessible through lattice QCD calculations. The electric charge chemical potential can be expanded as a series in terms of the $\mu_{\rm B}$, 
\begin{equation}
    {{\mu}_{\rm Q}}/{\mu}_{\rm B}=q_1 + q_3 ~\hat{\mu}_{\rm B}^2+\mathcal{O}(\hat{\mu}_{\rm B}^4),
\end{equation}
where the leading-order (LO) coefficient $q_1$ can be expressed as,
\begin{equation}
    q_1 =
\frac{
r\left( \chi_2^{\rm B}\chi_2^{\rm S} - \chi_{11}^{\rm BS}\chi_{11}^{\rm BS} \right)
-\left( \chi_{11}^{\rm BQ}\chi_2^{\rm S} -\chi_{11}^{\rm BS} \chi_{11}^{\rm QS} \right)
}{
\left( \chi_2^{\rm Q}\chi_2^{\rm S}  - \chi_{11}^{\rm QS} \chi_{11}^{\rm QS} \right)
- r \left(\chi_{11}^{\rm BQ}\chi_2^{\rm S} - \chi_{11}^{\rm BS}\chi_{11}^{\rm QS} \right)
},
\label{eq:q1}
\end{equation}
where $r = n_{\rm Q}/{n_{\rm B}}$ is the ratio of net electric charge to net baryon number density in the initial colliding nuclei. The next-to-leading order (NLO) coefficient expression can be found in Ref. \cite{Bazavov:2017dus}. 

\begin{figure}[!htp]
    \centering
    \includegraphics[width=0.40\textwidth]{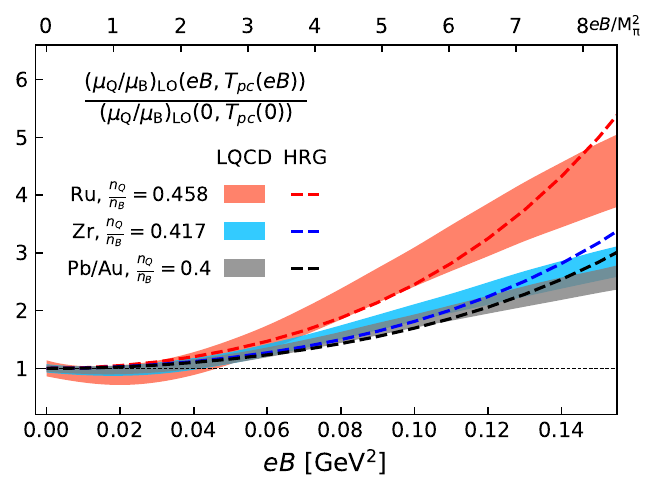}
    \includegraphics[width=0.42\textwidth]{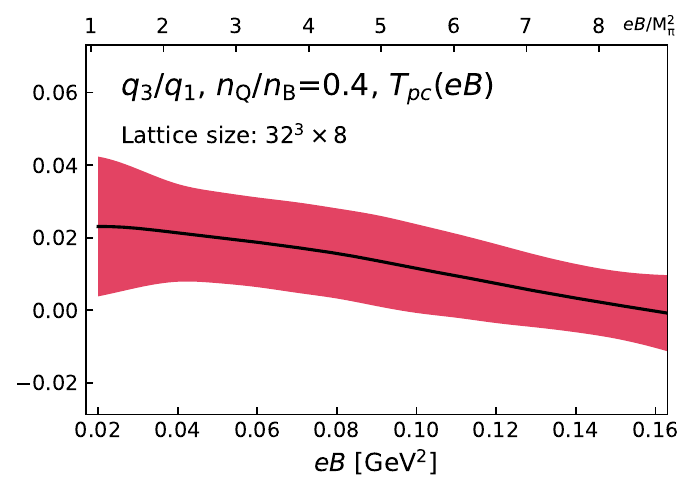}
    \caption{The left panel: the continuum estimated ratio of leading-order coefficient of ${{\mu}_{\rm Q}}/{\mu}_{\rm B}$ as a function of $eB$ at $T_{pc}$. The right panel: the $eB$ dependence of $q_3/q_1$ obtained on $32^3 \times 8$ lattices for $n_{\rm Q}/{n_{\rm B}} = 0.4$ at $T_{pc}$.}
    \label{fig:q1}
\end{figure}

The left panel of \autoref{fig:q1} illustrates the magnetic field dependence of the $R_{cp}$-like of $q_1$ for various colliding systems along the transition line. It can be seen that $q_1$ exhibits sensitivity not only to magnetic fields but also to the colliding nuclei involved in collisions. At $eB \sim 8 M_\pi^2$, although the charge-to-mass ratio of the Ruthenium (${}^{96}_{44}$Ru) collision system is only 10\% larger than that for Zirconium (${}^{96}_{40}$Zr), the ratio of $q_1$ in the ${}^{96}_{44}$Ru + ${}^{96}_{44}$Ru collision system exceeds that in the ${}^{96}_{40}$Zr + ${}^{96}_{40}$Zr by 50\%. The right panel shows the ratio of NLO coefficient $q_3$ to LO coefficient $q_1$ as a function of $eB$, calculated on lattices with $N_\tau=8$. The $q_3/q_1$ remains within approximately $5\%$. Furthermore, the discretization error for this quantity is minor \cite{Bazavov:2017dus,Bazavov:2012vg}, indicating that higher-order corrections to $\mu_{\rm Q}/\mu_{\rm B}$ in the continuum limit remain small. The pronounced enhancement under external $eB$ further suggests that $\mu_{\rm Q}/\mu_{\rm B}$ is well-suited to serve as a magnetometer in relativistic heavy ion collisions.  

\section{Conclusions}

In this work, we have presented the first lattice QCD computations of the second order fluctuations of conserved charges in an external magnetic field with a physical pion mass. The continuum estimates are based on $N_\tau =8$ and 12 lattices. We observe that both $\chi^{\rm BQ}_{11}$ and ${{\mu}_{\rm Q}}/{\mu}_{\rm B}$ are significantly influenced by $eB$, suggesting their potential utility as magnetometers for QCD. With the aid of the hadron resonance gas model, we can construct proxies for $\chi^{\rm BQ}_{11}$ using detectable hadrons in HIC experiments, and these proxies still exhibit the strong dependence of $\chi^{\rm BQ}_{11}$ on the magnetic field. To better align with the measurements in HIC experiments, future studies will incorporate kinematic cuts into the analysis of conserved charge fluctuations in the HRG framework \cite{Karsch:2015zna}. On the other hand, the ratio of ${{\mu}_{\rm Q}}/{\mu}_{\rm B}$ can also be obtained by thermal statistical fits to the particle yields experimentally. However, in such thermal fitting, an additional parameter $eB$ is required to account for the influence of the magnetic field on hadron energies. Finally, our results also provide a QCD benchmark for second order fluctuations of conserved charges for effective theories and model studies.

\acknowledgments
This work was supported in part by the National Natural Science Foundation of China under Grants No. 12293064, No. 12293060, and No. 12325508, and the National Key Research and Development Program of China under Contract No. 2022YFA1604900. The numerical simulations were performed using the GPU cluster at the Nuclear Science Computing Center of Central China Normal University (NSC3) and the Wuhan Supercomputing Center.

\bibliographystyle{JHEP.bst}
\bibliography{refs.bib}
\end{document}